\documentclass[twocolumn,english,aps,pra,superscriptaddress,showpacs,tightenlines]{revtex4-1}
\usepackage{amssymb}
\usepackage{amsmath}
\usepackage{amsfonts}
\usepackage{graphicx}
\usepackage{epstopdf} 
\usepackage{color}
\usepackage{txfonts}
\usepackage[normalem]{ulem}

\begin{document}
	
	\title{$\Lambda$-Type Giant Atom Mediated Controllable Single-Photon Transport in a One-Dimensional Chiral Waveguide
	}
	\author{Yi-Mei \surname{Wang}}
	\affiliation{Key Laboratory for Matter Microstructure and Function of Hunan Province, Hunan Research Center of the Basic Discipline for Quantum Effects and Quantum Technologies, Xiangjiang-Laboratory and Department of Physics, Hunan Normal University, Changsha 410081, China}
	
	\author{Jing \surname{Li}}
	\affiliation{School of Physics and Chemistry, Hunan First Normal University, Changsha 410205, China}
	
	\author{Lan \surname{Zhou} }
	\affiliation{Key Laboratory for Matter Microstructure and Function of Hunan Province, Hunan Research Center of the Basic Discipline for Quantum Effects and Quantum Technologies, Xiangjiang-Laboratory and Department of Physics, Hunan Normal University, Changsha 410081, China}
	
	\author{Jing \surname{Lu}}
	\affiliation{Key Laboratory for Matter Microstructure and Function of Hunan Province, Hunan Research Center of the Basic Discipline for Quantum Effects and Quantum Technologies, Xiangjiang-Laboratory and Department of Physics, Hunan Normal University, Changsha 410081, China}
	\affiliation{Institute of Interdisciplinary Studies, Hunan Normal University, Changsha, 410081, China
	}

	\begin{abstract}
		We investigate the single-photon scattering spectrum of a driven $\Lambda$-type giant atom system chirally coupled to a one-dimensional (1D) waveguide. By employing a real-space scattering approach, we obtain analytical solutions for the scattering amplitudes that remain valid in both Markovian and non-Markovian regimes. We observe that an external driving field induces a splitting of the transmission spectrum's dip into double dips, with the distance between the two dips increasing as the strength of the driving field increases. The chiral nature of the coupling allows for controlled switching between complete transmission and perfect reflection of incident photons. In the Markovian limit, we predict robust perfect transmission at specific phase values, independent of the driving field parameters.
		Moreover, in the non-Markovian regime, as the size of the giant atom increases, the oscillatory behavior of the scattering spectrum becomes more pronounced. Adjusting the giant atom size enables control over the number of decoupling points as well as the number of complete reflection points.

	\end{abstract}
	
	\date{\today}
	\maketitle

	
	\section{Introduction}
	The interaction between light and matter plays a crucial role in the fundamental sciences\cite{RevModPhys.91.025005,gutzler2021light}, supporting the rapid development of quantum technology. The light-matter interactions can be effectively realized across various physical platforms, including cavity quantum electrodynamics (QED) systems\cite{kimble1998strong,raimond2001manipulating,walther2006cavity}, circuit-QED systems\cite{blais2021circuit,blais2004cavity,wallraff2004strong} and waveguide-QED systems\cite{roy2017colloquium,gu2017microwave,trivedi2021optimal,sun2025cavity}. Unlike cavity-QED systems, waveguides typically support continuous modes, allowing for relaxation of the finite bandwidth of the field\cite{roy2017colloquium}. Especially when multiple atoms are coupled to a one-dimensional (1D) waveguide, the coherent and dissipative interactions between the emitters can be mediated through the propagation modes, leading to many interesting phenomena, including superradiance and subradiance\cite{dicke1954coherence,vetter2016single,van2013photon,zhang2019theory,ke2019inelastic,wang2020controllable,PhysRevResearch.2.043149,dinc2019exact}, long-range entanglement between distant emitters\cite{zheng2013persistent,gonzalez2014generation,facchi2016bound,mirza2016multiqubit}, the formation of photonic band gaps\cite{fang2015waveguide,greenberg2021waveguide,he2021atomic}, cavity quantum electrodynamics with atomic mirrors\cite{chang2012cavity,mirhosseini2019cavity}, topologically protected spectroscopy\cite{nie2021topology}, topologically enhanced nonreciprocal scattering\cite{PhysRevApplied.15.044041}, asymmetric Fano line shapes\cite{tsoi2008quantum,cheng2012fano,liao2015single,cheng2017waveguide,mukhopadhyay2019multiple}, and electromagnetically induced transparency (EIT) without a control field\cite{shen2007stopping,fang2017multiple,mukhopadhyay2020transparency}. These studies are typically based on point-like atoms, where the wavelength of light is usually much larger than the size of the point-like atoms. Therefore, the dipole approximation can be employed to describe the interaction between light and atoms\cite{walls2008light}.
	
	Recently, with the development of modern nanotechnology, a new type of waveguide-QED structure has attracted widespread attention\cite{frisk2014designing,kockum2021quantum}, in which the quantum emitters are so-called giant atoms (GAs). Unlike small quantum emitters, due to the large size of giant atoms, they couple to the waveguide at multiple coupling points, providing new channels for quantum interference. Many interesting quantum optical effects induced by giant atoms have been predicted, including frequency-dependent relaxation rate and Lamb shifts\cite{frisk2014designing,kannan2020waveguide,vadiraj2021engineering}, non-Markovian dynamics\cite{xu2024catch,lim2023oscillating,yin2022non}, the formation of bound states\cite{sun2025cavity,lim2023oscillating,guo2020oscillating,zhao2020single,wang2021tunable,he2024emergent,yang2025coherent,lim2023oscillating}, and chiral light-matter interactions\cite{soro2022chiral,du2022giant,li2024single}. It is particularly noteworthy that most current studies focus on two-level giant atom systems. While these have revealed many new phenomena, the relatively simple energy level structure limits the realization of more complex quantum effects. 
	
	Based on this, we propose a one-dimensional waveguide chirally coupled to a $\Lambda$-type three-level giant atom system. Using the real-space scattering approach, we derive a general analytical expression for the single-photon scattering amplitude in the $\Lambda$-type giant atom system. Based on these general expressions, we further analyzed three important phenomena: 1) the induction of spectral splitting by tuning the driving field $\Omega$, which cannot be observed in two-level atomic systems; 2) the ability to achieve a complete transition between full transmission and full reflection of photons by varying the coupling strength. This characteristic is also unattainable in non-chiral environments; and 3) the size of giant atoms can regulate the number of decoupling points and the complete reflection points.
	
	The paper is organized as follows: In Sec.~\ref{Sec:2}, we outline the theoretical model of the system by
	presenting the Hamiltonian. We give the calculation method and solution to the model. In Sec.~\ref{Sec:3},
	the transmission properties of single photons are analyzed under the Markovian regime and non-Markovian regime, respectively, with a focus on exploring the effects of chiral coupling strength and the driving field on their behavior. Finally, we
	summarize the single-photon transport and give the conclusions in Sec.~\ref{Sec:4}.

	\section{\label{Sec:2} Model Hamiltonian}
	\begin{figure}[tbp]
		\includegraphics[width=0.5\textwidth]{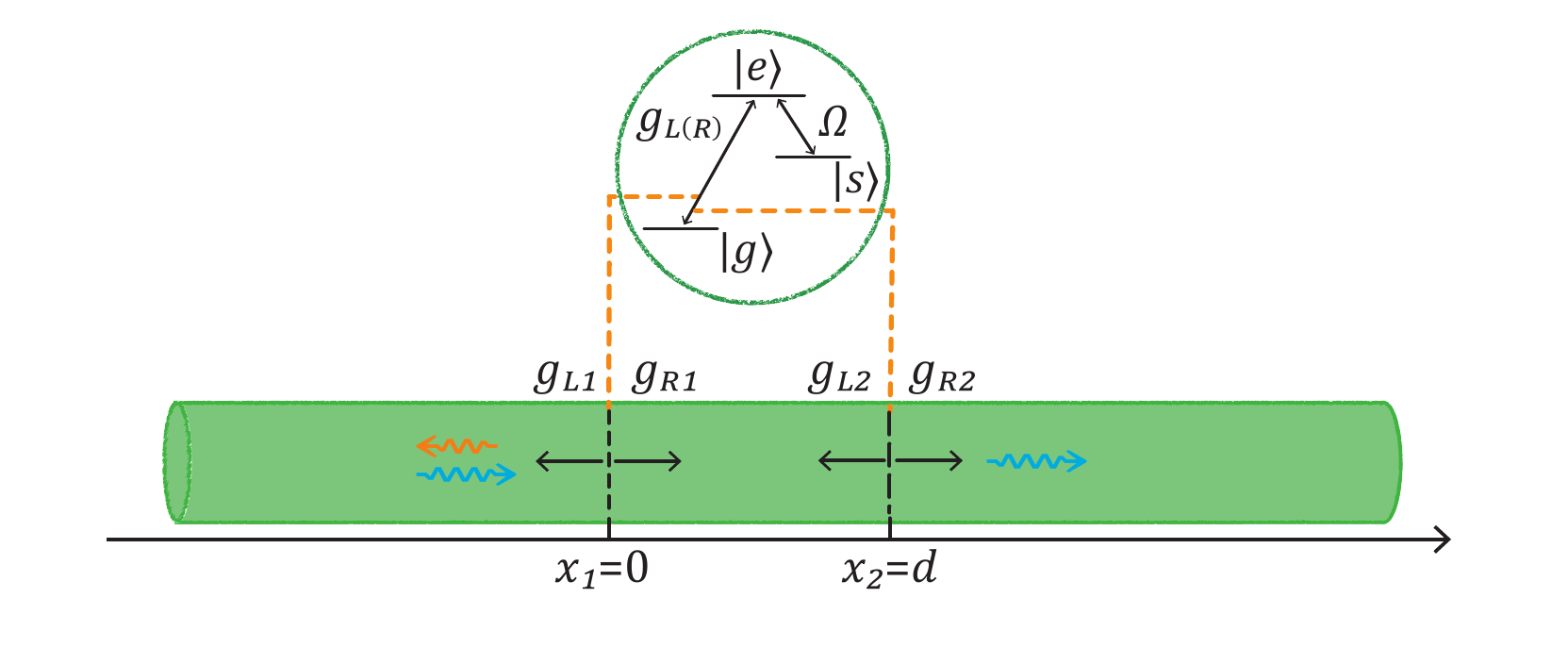}
		\caption{Schematic of the giant waveguide-QED system. A $\Lambda$-type three-level giant atom is coupled to a one-dimensional infinitely long waveguide at positions $x_1=0$ and $x_2=d$. The transition between the excited state $\left\vert e\right\rangle$ and ground state $\left\vert g\right\rangle$ is chirally coupled to the waveguide at both coupling positions, while the metastable state $\left\vert s\right\rangle$ is decoupled from the waveguide but driven by an external field with frequency $\Omega$. }
		\label{fig1}
	\end{figure}
	We consider a $\Lambda$-type three-level giant atom driven by an external field and chirally coupled to a one-dimensional (1D) waveguide at two discrete points, as schematically shown in Fig.~\ref{fig1}. For simplicity, hereafter we assume that the distance between two coupling points is $d$. In the rotating-wave approximation, the Hamiltonian of the total system reads ($\hbar=1$)
	\begin{equation}  \label{1-1}
		\hat{H}=\hat{H}_{a}+\hat{H}_{\omega}+\hat{H}_{int},
	\end{equation}%
	where $\hat{H}_{a}$ and $\hat{H}_{\omega}$ are the free Hamiltonians of the $\Lambda$-type three-level giant atom and the waveguide, respectively. $\hat{H}_{int}$ describes the interaction between the giant atom and the waveguide.
	
	The Hamiltonian of the $\Lambda$-type three-level giant atom takes the form
	\begin{equation}  \label{1-2}
		\hat{H}_{a}=\omega _{e}\left\vert e\right\rangle \left\langle e\right\vert
		+\omega _{s}\left\vert s\right\rangle \left\langle s\right\vert +\Omega
		\left( \left\vert s\right\rangle \left\langle e\right\vert +\left\vert
		e\right\rangle \left\langle s\right\vert \right),
	\end{equation}%
	with $\omega _{e}$ and $\omega _{s}$ the energies of the excited state $\left\vert e\right\rangle$ and the metastable state $\left\vert s\right\rangle$ with respect to the ground state $\left\vert g\right\rangle$, respectively. The transition between states $\left\vert e\right\rangle$ and $\left\vert s\right\rangle$ is driven by a classical field with driving strength $\Omega$.
	
	The Hamiltonian of the waveguide is given by
	\begin{eqnarray}  \label{1-3}
		\hat{H}_{\omega } &=&\int_{-\infty }^{+\infty }dx\hat{a}_{L}^{\dag }\left(
		x\right) \left( \omega _{0}+iv_{g}\frac{\partial }{\partial x}\right) \hat{a}%
		_{L}\left( x\right)  \\ \nonumber
		&&+\int_{-\infty }^{+\infty }dx\hat{a}_{R}^{\dag }\left( x\right) \left(
		\omega _{0}-iv_{g}\frac{\partial }{\partial x}\right) \hat{a}_{R}\left(
		x\right),
	\end{eqnarray}
	where $\hat{a}_{R}^{\dag }\left( x\right)$ ($\hat{a}_{L}^{\dag }\left( x\right)$) is the creation operators of the right-propagating (left-propagating) mode at position $x$ and $v_{g}$ is the group velocity. $\hat{a}_{R}\left( x\right)$ ($\hat{a}_{L}\left( x\right)$) is its adjoint operator. $\omega_{0}$ is a chosen frequency that is away from the cutoff frequency, around which the dispersion relation of the waveguide can be linearized as $\omega_{k}=\omega_{0}+v_{g}(k-k_{0})$. 
	
	The Hamiltonian of the interaction between the giant atom and the waveguide can be written as
	\begin{eqnarray}  \label{1-4}
		\hat{H}_{int} &=&\int_{-\infty }^{+\infty }dx\sum_{j=1}^{2}\delta \left(
		x-x_{j}\right) \sqrt{v_{g}}e^{-ik_{0}x}g_{Rj}\hat{a}_{R}^{\dag }\left(
		x\right) \left\vert g\right\rangle \left\langle e\right\vert  \\ \nonumber
		&&+\int_{-\infty }^{+\infty }dx\sum_{j=1}^{2}\delta \left( x-x_{j}\right)
		\sqrt{v_{g}}e^{ik_{0}x}g_{Lj}\hat{a}_{L}^{\dag }\left( x\right) \left\vert
		g\right\rangle \left\langle e\right\vert +H.c.,
	\end{eqnarray}%
	with $g_{Rj}$ and $g_{Lj}$ being the coupling strengths for the giant atom interacting with a right-going and left-going photon at the position $x=x_{j}$. Here, we consider the coupling strengths to be real.
	In the case of a single excitation, the state vector of the system can be written as
	\begin{eqnarray}  \label{1-5}
		\left\vert \psi \right\rangle &=&\int_{-\infty }^{+\infty }dx\left[
		C_{gL}\left( x\right) a_{L}^{\dag }\left( x\right) +C_{gR}\left( x\right)
		a_{R}^{\dag }\left( x\right) \right] \left\vert g,0\right\rangle \\ \nonumber
		&&+C_{e}\left\vert e,0\right\rangle +C_{s}\left\vert s,0\right\rangle,
	\end{eqnarray}%
	where $C_{gL}\left( x\right)$ ($C_{gR}\left( x\right)$) is the probability amplitude of generating a left-propagating (right-propagating) photon at the waveguide $x$. $C_{e}$ is the probability amplitude for the atom to be in the excited state and for there to be no photons in the waveguide. $C_{s}$ is the probability amplitude for the atom to be in the metastable state and for there to be no photons in the waveguide. Solving the stationary Schr$\ddot{o}$dinger equation $\hat{H}\left\vert \psi \right\rangle=E\left\vert \psi \right\rangle$ with Eqs.~(\ref{1-1}) and (\ref{1-5}), the coupled equations can be written
	\begin{eqnarray}\label{1-6}
		E C_{gL}(x) &=& C_{gL}(x)\left(\omega_0 + i v_{g} \frac{\partial}{\partial x}\right) +\sqrt{v_{g}}C_{e}\sum_{j=1}^{2}e^{ik_{0}x}g_{Lj}\delta \left(x-x_{j}\right), \nonumber\\
		E C_{gR}(x) &=& C_{gR}(x)\left(\omega_0 - i v_{g} \frac{\partial}{\partial x}\right) +\sqrt{v_{g}}C_{e}\sum_{j=1}^{2}e^{-ik_{0}x}g_{Rj}\delta \left(x-x_{j}\right), \nonumber\\ 
		(E-\omega_{e}) C_{e}&=& \Omega C_{s}+\sqrt{v_{g}} \sum_{j=1}^{2} \left[ C_{gR}(x_{j}) e^{i k_0 x_{j}} g_{Rj}^{*} + C_{gL}(x_{j}) e^{-i k_0 x_{j}} g_{Lj}^{*} \right],\nonumber\\
		(E-\omega _{s}) C_{s}&=&\Omega C_{e}.
	\end{eqnarray}
	We consider that a single photon with energy $E$ is initially injected from the left-hand side of the waveguide. In the regimes of $x<x_{1}$, $x_{1}<x<x_{2}$, and $x>x_{2}$, the amplitude $C_{gR}$ and  $C_{gL}$ can be written in the form
	\begin{eqnarray} 
	C_{gR}\left( x\right)  &=&\left\{
	\begin{array}{c}
		e^{i\left( k-k_{0}\right) x},x<x_{1} \\
		Ae^{i\left( k-k_{0}\right) x},x_{1}<x<x_{2} \\
		te^{i\left( k-k_{0}\right) x},x>x_{2}%
	\end{array}%
	\right., \nonumber\\
	C_{gL}\left( x\right)  &=&\left\{
	\begin{array}{c}
		re^{-i\left( k-k_{0}\right) x},x<x_{1} \\
		Be^{-i\left( k-k_{0}\right) x},x_{1}<x<x_{2}%
	\end{array}%
	\right..\label{1-7}
	\end{eqnarray}
	Here, $r$ and $t$ are the single photon reflection and transmission amplitudes, respectively. Combine Eq.~(\ref{1-6}) and Eq.~(\ref{1-7}), we obtain the reflection and transmission amplitudes
	\begin{widetext}
	\begin{eqnarray}
			t&=& \frac{
				\Delta - \frac{\Omega^2}{\Delta + \omega_{e} - \omega_{s}} 
				- \sum_{i,j=1}^2 \sqrt{\gamma_{Ri}\gamma_{Rj}} \sin\phi_{ij} 
				- \sum_{i,j=1}^2 \sqrt{\gamma_{Li}\gamma_{Lj}} \sin\phi_{ij}
				+ i \biggl(
				\sum_{i,j=1}^2 \sqrt{\gamma_{Li}\gamma_{Lj}} \cos\phi_{ij}
				- \sum_{i,j=1}^2 \sqrt{\gamma_{Ri}\gamma_{Rj}} \cos\phi_{ij}
				\biggr)	}{
				\Delta - \frac{\Omega^2}{\Delta + \omega_{e} - \omega_{s}}
				- \sum_{i,j=1}^2 \sqrt{\gamma_{Ri}\gamma_{Rj}} \sin\phi_{ij}
				- \sum_{i,j=1}^2 \sqrt{\gamma_{Li}\gamma_{Lj}} \sin\phi_{ij}
				+ i \biggl(
				\sum_{i,j=1}^2 \sqrt{\gamma_{Li}\gamma_{Lj}} \cos\phi_{ij}
				+ \sum_{i,j=1}^2 \sqrt{\gamma_{Ri}\gamma_{Rj}} \cos\phi_{ij}
				\biggr)},\label{1-9}\\
				r &=& -\frac{
					i \sum_{i,j=1}^2 2\sqrt{\gamma_{Li}\gamma_{Rj}} e^{ik(x_i + x_j)}
				}{
					\Delta - \frac{\Omega^2}{\Delta + \omega_{e} - \omega_{s}}
					- \sum_{i,j=1}^2 \sqrt{\gamma_{Ri}\gamma_{Rj}} \sin\phi_{ij}
					- \sum_{i,j=1}^2 \sqrt{\gamma_{Li}\gamma_{Lj}} \sin\phi_{ij}
					+ i \biggl(
					\sum_{i,j=1}^2 \sqrt{\gamma_{Li}\gamma_{Lj}} \cos\phi_{ij}
					+ \sum_{i,j=1}^2 \sqrt{\gamma_{Ri}\gamma_{Rj}} \cos\phi_{ij}
					\biggr)},\label{1-10}								
	\end{eqnarray}
	\end{widetext}
	where $\Delta = \omega - \omega_{e}$ represents the detuning between the incident photon frequency and the atomic transition frequency;
	Furthermore, we define the decay rate of the j-th coupling point as $\gamma_j = \gamma_{Lj} + \gamma_{Rj}$, where $\gamma_{Lj} = g_{Lj}^2/2$ and $\gamma_{Rj} = g_{Rj}^2/2$ represent the left and right decay contributions, respectively. 
	The phase difference between coupling points is given by $\phi_{ij} = |i-j|\phi$, where $\phi = (\Delta + \omega_{e})\tau = \Delta\tau + \theta$ represents the accumulated phase between two adjacent coupling points, with $\tau$ being the propagation time between them.
	Neglecting atomic dissipation, the reflection and transmission coefficients satisfy the relation $1=R+T$, where $R=|r|^{2}$ is the reflection coefficient and $T=|t|^{2}$ is the transmission coefficient.
	
	\section{\label{Sec:3} Single photon scattering}
	
    In the following discussion, we consider identical decay rates at both coupling points (i.e., $\gamma_{1} = \gamma_{2}$). The chiral coupling
    can be divided into two distinct cases: (1) bidirectional
	even coupling (BEC) $\gamma_{L1}=\gamma_{L2},\; \gamma_{R1}=\gamma_{R2},$ but $\gamma_{L1}\neq\gamma_{R1}$; (2) bidirectional uneven coupling (BUEC) $\gamma_{L1}=\gamma_{R2},\; \gamma_{R1}=\gamma_{L2}$.
	Throughout this work, we focus on the resonant condition $\omega_s = \omega_e$, under which the transmission probability takes the form
	\begin{widetext}
	\begin{eqnarray}
		T=\frac{
			[\Delta-2(\sqrt{\gamma_{L1}\gamma_{L2}}+\sqrt{\gamma_{R1}\gamma_{R2}})\sin(\Delta\tau+\theta)-\frac{\Omega^{2}}{\Delta}]^{2}+
			[\gamma_{L1}+\gamma_{L2}-\gamma_{R1}-\gamma_{R2}+2(\sqrt{\gamma_{L1}\gamma_{L2}}-\sqrt{\gamma_{R1}\gamma_{R2}})\cos(\Delta\tau+\theta)]^{2}}
			{[\Delta-2(\sqrt{\gamma_{L1}\gamma_{L2}}+\sqrt{\gamma_{R1}\gamma_{R2}})\sin(\Delta\tau+\theta)-\frac{\Omega^{2}}{\Delta}]^{2}+
			[\gamma_{L1}+\gamma_{L2}+\gamma_{R1}+\gamma_{R2}+2(\sqrt{\gamma_{L1}\gamma_{L2}}+\sqrt{\gamma_{R1}\gamma_{R2}})\cos(\Delta\tau+\theta)]^{2}
			}.\label{1-11}
	\end{eqnarray}
	\end{widetext}
	From the above expression, it can be observed that when $\Delta\tau+\theta=(2n+1)\pi$ ($n\in Z$), $\gamma_{L1}=\gamma_{L2}$ and $\gamma_{R1}=\gamma_{R2}$ (BEC), 
	although the transmission probability consistently remains unity, the denominator in Eq.~(\ref{1-11}) develops singularities at the detuning points $\Delta=\pm\Omega$. This physical scenario indicates a complete decoupling between the giant atom and waveguide, leading to perfect transmission of incident photons with unit probability. However, under the identical phase condition with $\gamma_{L1}=\gamma_{R2}$ and $\gamma_{R1}=\gamma_{L2}$ (BUEC), Eq.~(\ref{1-11}) reduces to
	\begin{eqnarray}
		T=\frac{(\Delta-\frac{\Omega^{2}}{\Delta})^2}{(\Delta-\frac{\Omega^{2}}{\Delta})^2+[(\sqrt{\gamma_{L1}}-\sqrt{\gamma_{L2}})^2+(\sqrt{\gamma_{R1}}-\sqrt{\gamma_{R2}})^2]^2}.
	\end{eqnarray}
	Here, when $\Delta=\pm\Omega$, the photon undergoes total reflection.
	In other words, an incident photon can be totally reflected or totally transmitted for the phases $\Delta\tau + \theta = (2n+1)\pi$ when the chirality of couplings is tuned between the BEC regime and the BUEC regime.
	
	Next, we investigate single-photon scattering in both Markovian and non-Markovian regimes.
	The system exhibits Markovian dynamics when $\tau\ll1/\tilde{\Gamma}$($\tilde{\Gamma}=\sum_{j=1}^{2} \gamma_{j}$), while it transitions to non-Markovian behavior occur at $\tau\simeq 1/{\tilde{\Gamma}}$, where retardation effects become significant in the giant atom-waveguide coupling.
	
	\subsection{The markovian regime}

	In the subsection, we focus on single-photon scattering in the Markovian regime. At this regime, the propagation time satisfies the condition $\tau\tilde{\Gamma}\ll1$, and hence the detuning-dependent terms $\Delta\tau$ in Eq.~(\ref{1-11}) can be ignored, such that $(\theta+\Delta\tau) \approx \theta$. By changing the distance between adjacent coupling points $d$ and the transition frequency of the giant atom $\omega_{e}$, the phase $\theta$ can be valued in the range $\theta \in \left[ 0,2\pi \right] $. 	
	One can observe that single-photon scattering results in either total reflection or total transmission, depending on the coupling strength and the phase $\theta$. We point out that when $\Omega=0$, the three-level giant atom degenerates into a two-level system. 
	\begin{figure*}[tbp]
		\begin{center}
			\includegraphics[width=16cm]{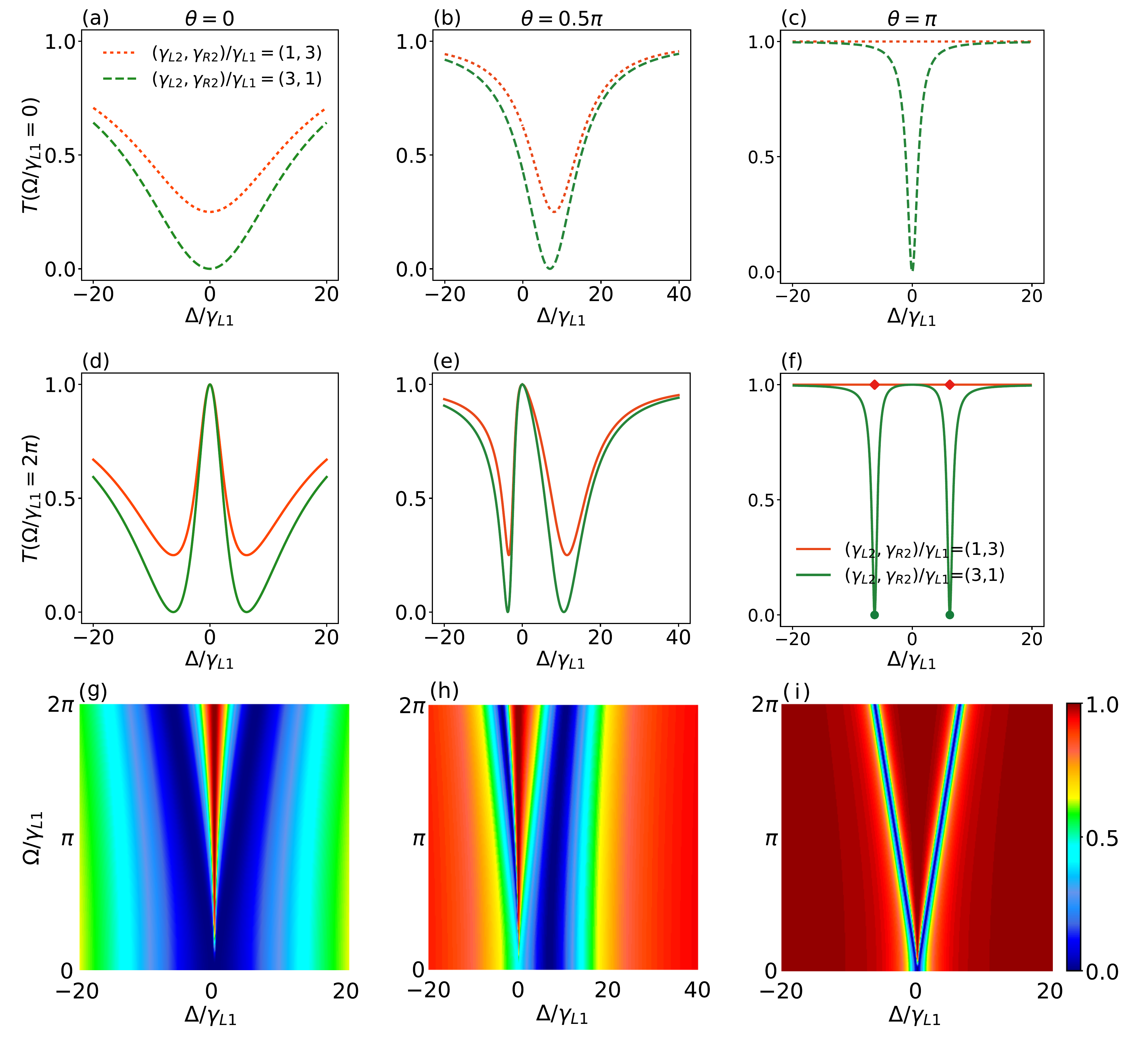}
			\caption{The driving field modulates the transmission probabilities $T$ in the Markovian regime. (a)-(f) Transmission probabilities versus $\Delta$ for different $\theta$ and $\Omega$. The solid and dashed lines represent driving fields with amplitudes of $\Omega/\gamma_{L1}=2\pi$ and $0$, respectively. The left and right decay rates exhibit ratios of $(\gamma_{L1}, \gamma_{R1})/\gamma_{L1} = (1, 3)$ for the first coupling point, and $(\gamma_{L2}, \gamma_{R2})/\gamma_{L1} =\{(1, 3), (3, 1)\}$ for the second coupling point, corresponding to the red line and green line. (a,d): $\theta= 0$; (b,e): $\theta= 0.5\pi$; (c,f): $\theta=\pi$. (g)-(i): Transmission probabilities versus $\Delta$ and $\Omega$ with $(\gamma_{L1}, \gamma_{R1},\gamma_{L2}, \gamma_{R2})/\gamma_{L1} = (1,3,3,1)$. (g): $\theta= 0$; (h): $\theta= 0.5\pi$; (i): $\theta=\pi$.}
			\label{fig2}
		\end{center}
	\end{figure*}

	
	In Fig.~\ref{fig2}(a-f), we plot the transmission spectrum of single-photon scattering as a function of the detuning $\Delta$. We systematically examine how the phase $\theta$ and coupling strength affect the single-photon transmission spectrum, comparing scenarios with and without an external drive field. 
	When $\Omega=0$ and $\theta=0$, the transmission probability of photons at the resonant frequency point $\Delta=0$ reaches its minimum. By adjusting the coupling strength to satisfy $\gamma_{L1}=\gamma_{R2},\; \gamma_{R1}=\gamma_{L2}$, the transmission probability can be further suppressed to zero, as shown by the green dashed line in Fig.~\ref{fig2}(a).
	When $\theta =\pi/2 $, the dip in transmission probability shifts to $\Delta = 2\left(\sqrt{\gamma_{L1}\gamma_{L2}} + \sqrt{\gamma_{R1}\gamma_{R2}}\right)$. Especially for $\gamma_{L1}+\gamma_{L2}=\gamma_{R1}+\gamma_{R2}$, transmission probability $T=0$, as shown by the green dashed line in Fig.~\ref{fig2}(b).
	Interestingly, as shown in Fig.~\ref{fig2}(c), when $\theta=\pi$, the full reflection point of the resonance frequency can be realized to become the full transmission point by adjusting the coupling strength. For example, if $\gamma_{L1}=\gamma_{L2}$ and $ \gamma_{R1}=\gamma_{R2}$ (BEC), the transmission coefficient reaches unity ($T=1$), as indicated by the red dash line. Conversely, when $|\sqrt{\gamma_{L1}}-\sqrt{\gamma_{L2}}|=|\sqrt{\gamma_{R1}}-\sqrt{\gamma_{R2}}|$, perfect reflection ($T = 0$) occurs at $\Delta = 0$, as indicated by the green dashed curve.
	These results are consistent with the results of Reference\cite{li2024single}. But there's only one frequency point where one can get full reflection or transmission.
	
	However, if $\Omega\neq 0$, specific phase conditions enable the regulation of chiral coupling to achieve either total reflection or total transmission of single photons at two distinct frequency points. When $\theta=0$ and $\gamma_{L1}=\gamma_{R2},\; \gamma_{R1}=\gamma_{L2}$, there are two frequency points ($\Delta=\pm\Omega$) where the total reflection $T=0$ of photons can be achieved, as shown by the green solid line in Fig.~\ref{fig2}(d). It should be noted that there is a peak in the transmission spectrum at $\Delta=0$, but this phenomenon does not exist in Fig.~\ref{fig2}(a). The transmission spectrum exhibits two dips of equal width, displaying a symmetric lineshape centered at $\Delta=0$. However, when $\theta=\pi/2$, the symmetry of the transmission spectrum is broken, as shown in Fig.~\ref{fig2}(e). 
	When $\theta=\pi$, it can be observed that in the BEC regime the transmission probability can be expressed as $T \equiv 1$, indicating that the transmission probability remains unity regardless of the driving field, as shown by the red solid line in the Fig.~\ref{fig2}(f). However, in the BUEC region, two frequency points exhibit total photon reflection, as shown by the green solid line in Fig.~\ref{fig2}(f).
	Remarkably, we observe two fundamentally distinct behaviors at $\Delta=\pm\Omega$ for different chiral coupling configurations. As previously discussed for the phase condition $\theta=(2n+1)\pi$, the transmission characteristics reveal a striking contrast: for BEC regime, singularities emerge in the denominator of Eq.~(\ref{1-11}) at these detuning frequencies, indicating complete giant atom-waveguide decoupling that manifests as the two red diamonds in Fig.~\ref{fig2}(f). Conversely, for BUEC regime under identical phase conditions, the transmission probability vanishes entirely at $\Delta=\pm\Omega$, corresponding to the two green points in the same figure, demonstrating how coupling asymmetry can dramatically alter the system's response at these critical frequencies.
	Moreover, as evident from the figure, the scattering spectrum maintains $\Delta=0$ symmetry when $\theta=n\pi$, whether or not a driving field is present. This symmetry is broken for $\theta \neq n\pi$.

	Next, in the BUEC regime: $(\gamma_{L2}, \gamma_{R2})/\gamma_{L1} = (3, 1)$, we plot in Fig.~\ref{fig2}(g-i) the transmission probability $T$ versus the detuning $\Delta$ and the driving field $\Omega$ in three cases, i.e, (g) $\theta=0$, (h) $\theta=0.5\pi$ and (i) $\theta=\pi$.
	As shown in Fig.~\ref{fig2}(g-i), a pronounced transmission dip emerges at $\Omega \approx 0$, independent of $\theta$ values. In particular, for $\theta = n\pi$, photons are completely reflected at the resonance point $T(\Delta=0)=0$. Under finite driving ($\Omega \neq 0$), the transmission spectrum evolves from a single-dip profile to a double-dip structure. A larger $\Omega$ results in a greater separation between the two transmission dips, and vice versa. 
	For the BEC regime: $(\gamma_{L2}, \gamma_{R2})/\gamma_{L1} = (1, 3)$, when $\theta = (2n+1)\pi$, as previously mentioned, the transmission probability remains unity ($T \equiv 1$) regardless of the driving field. For other values of $\theta$, the transmission probability $T$, plotted as a function of both detuning $\Delta$ and the driving field $\Omega$, shows behavior similar to that in the BUEC regime.
	
	\subsection{The non-Markovian regime}
	\begin{figure}[tbp]
		\includegraphics[width=0.45 \textwidth]{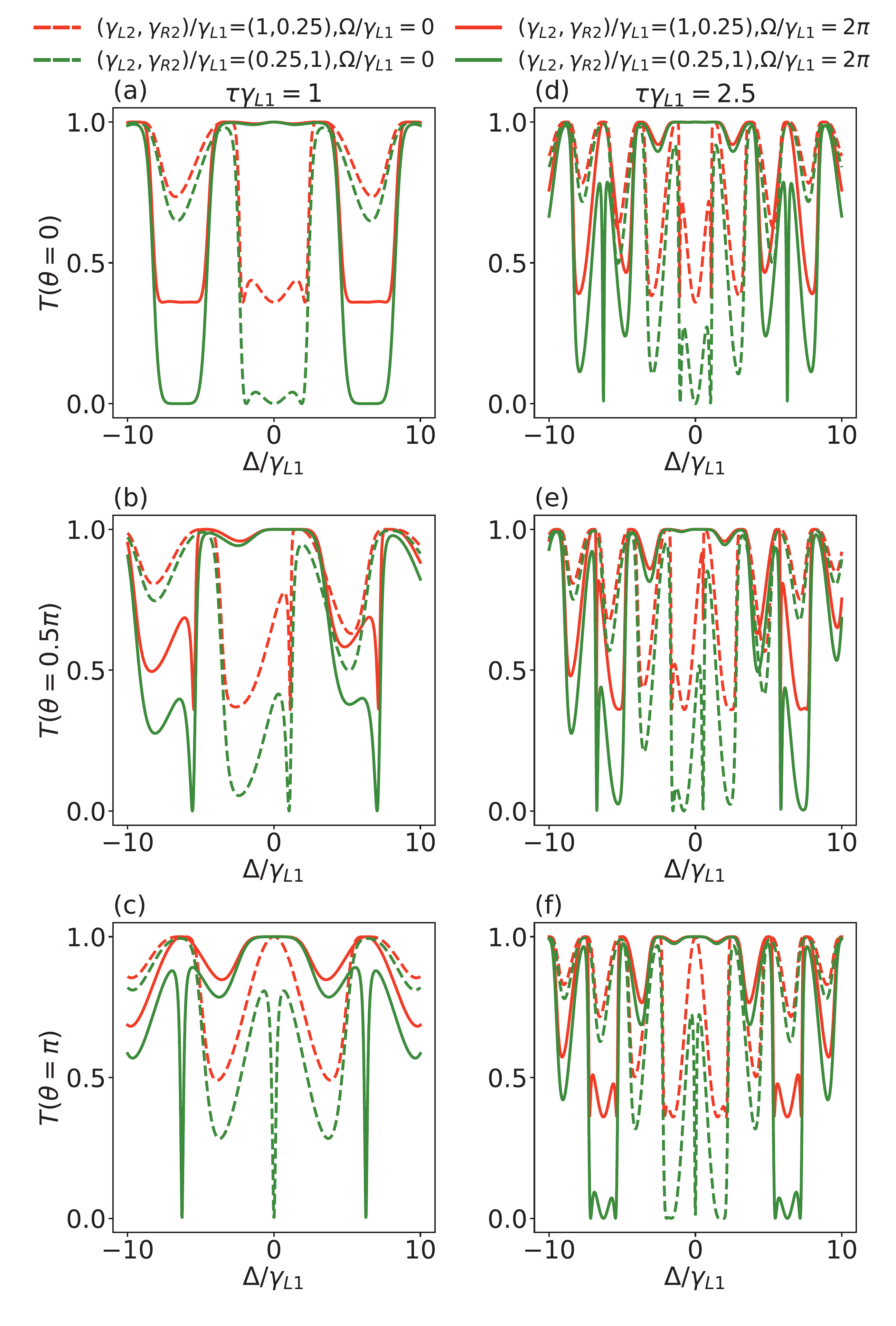}
		\caption{
			The transmission probability $T$ in the non-Markovian regime. The coupling ratios are fixed at $(\gamma_{L1}, \gamma_{R1})/\gamma_{L1} = (1, 0.25)$ for the first coupling point, while the second coupling point takes $(\gamma_{L2}, \gamma_{R2})/\gamma_{L1} =(1, 0.25)$ for the BEC regime (red curves) and $(\gamma_{L2}, \gamma_{R2})/\gamma_{L1} =(0.25, 1)$ for the BUEC regime (green curves). Solid and dashed lines represent finite ($\Omega/\gamma_{L1} = 2\pi$) and zero driving amplitudes, respectively. Panels (a)-(c): $\tau \gamma_{L1} = 1$ for $\theta= 0,0.5\pi,\pi$; (d)-(f): $\tau \gamma_{L1} = 2.5$ for the same $\theta$ values. } 
		\label{fig4}
	\end{figure} 
	So far, our analysis has been restricted to the Markovian regime. Now we consider the non-Markovian regime which the propagation time $\tau$ is comparable to or greater than the characteristic relaxation $1/\tilde{\Gamma}$. In the non-Markovian regime, the accumulated phase of photons between adjacent coupling points in the waveguide is given by $\theta+\Delta\tau$.
	
	In Fig.~\ref{fig4}, we plot the transmission probability $T$ as a function of detuning $\Delta$ for fixed values of the dimensionless time delay $\tau\gamma_{L1}$ and phase $\theta$.
	We begin by considering the non-Markovian regime with $\tau \gamma_{L1} = 1$, as illustrated in Fig.~\ref{fig4}(a-c). 
	When $\theta = 0$ and $\Omega = 0$, the transmission spectrum shows three distinct dips near $\Delta = 0$.
	Notably, in the BUEC regime $(\gamma_{L2}, \gamma_{R2})/\gamma_{L1} = (0.25, 1)$, all three transmission dips attain perfect suppression ($T=0$), as shown by the green dashed line in Fig.~\ref{fig4}(a). However, when $\Omega/\gamma_{L1}=2\pi$, one can readily observe a wide spectral range spanning both sides of $\Delta = 0$ that exhibits invariant transmission probability.
	Especially for $(\gamma_{L2}, \gamma_{R2})/\gamma_{L1} = (0.25, 1)$ (BUEC), the transmission probability is maintained around zero, as clearly demonstrated by the green solid line in Fig.~\ref{fig4}(a). 
	When $\theta = 0.5\pi$, the displacement of the dip arises, which breaks the symmetry of the transmission spectrum, as shown in Fig.~\ref{fig4}(b). In Fig.~\ref{fig4}(c), when $\theta=\pi$ and $\Omega=0$, the transmission probability exhibits oscillations on both sides of $\Delta=0$. When $\Omega/\gamma_{L1}=2\pi$, a wider region of perfect transmission is formed around $\Delta=0$. Furthermore, when $\tau \gamma_{L1}=2.5$, we observe pronounced oscillations in the transmission spectrum, which persist regardless of chiral coupling or the presence of driving fields, as shown in Fig.~\ref{fig4}(d-f). These oscillations are accompanied by the emergence of additional side valleys adjacent to the primary minima. These observations collectively demonstrate that in non-Markovian regime, as the non-Markov property increases, the transmission probability exhibits progressively stronger oscillations.
	
	\begin{figure*}[tbh]
		\begin{center}
			\includegraphics[width=16cm]{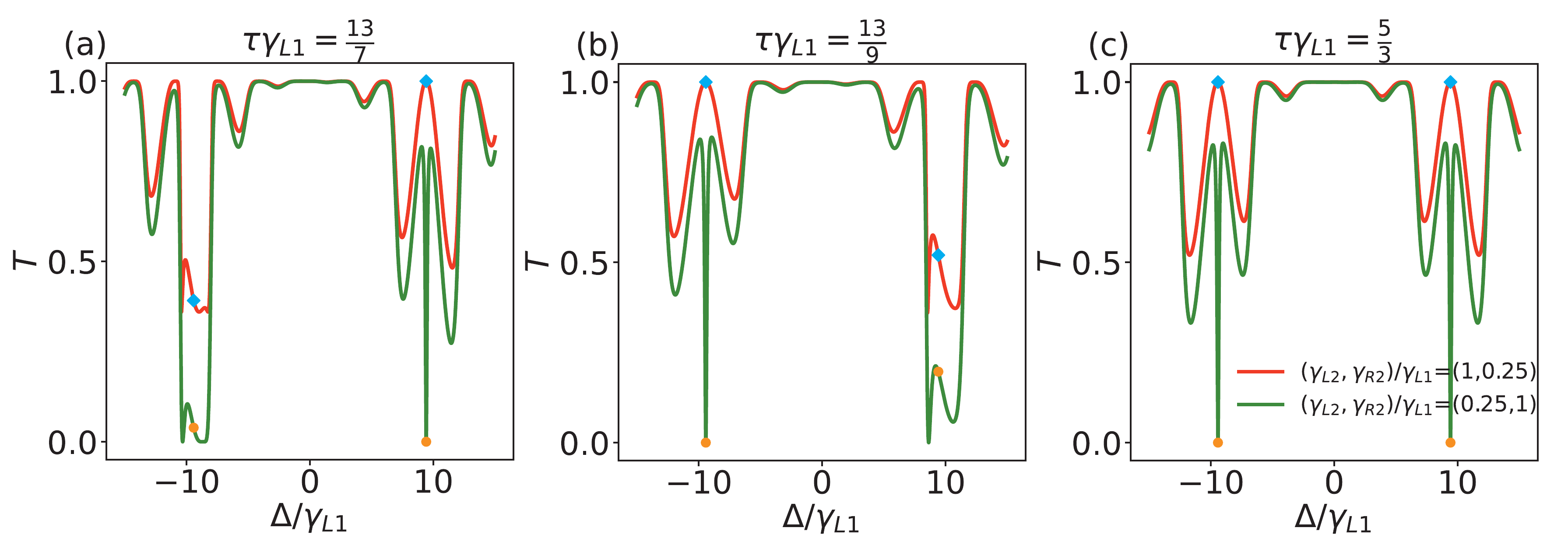}
			\caption{Transmission probability versus $\Delta$ for different coupling distance. We set the normalized decay rates of the first coupling point as $(\gamma_{L1}, \gamma_{R1})/\gamma_{L1} = (1, 0.25)$, while for the second coupling point we employ $(\gamma_{L1}, \gamma_{R1})/\gamma_{L1} = (1, 0.25)$ for the BEC regime (red curves) and $(\gamma_{L2}, \gamma_{R2})/\gamma_{L1} =(0.25, 1)$ for the BUEC regime (green curves), with the system parameters fixed at $\omega_e/\gamma_{L1} = 3000\pi$ and the driving amplitude $\Omega/\gamma_{L1} =3\pi$.
			The blue diamonds indicate complete decoupling in the BEC regime at $\Delta=\pm\Omega$, while yellow points mark perfect reflection in the BUEC regime at the same detunings. (a) $\tau\gamma_{L1}=13/7$; (b) $\tau\gamma_{L1}=13/9$; (c) $\tau\gamma_{L1}=5/3$.}
			\label{fig5}
		\end{center}
	\end{figure*}

	Under the Markovian limit, perfect transmission or reflection can be achieved by tuning the chiral coupling when $\theta = (2n + 1)\pi$. Similarly, in the non-Markovian regime, perfect transmission or perfect reflection can also be realized.
	When $\theta + \Delta\tau = (2n + 1)\pi$, the perfect transmission condition occurs in the BEC regime, while complete reflection occurs in the BUEC regime. When $\Delta=\pm\Omega$, the complete transmission indicates a decoupling between the giant atom and the waveguide. 
	Whereas in the BUEC regime, at $\Delta = \pm \Omega$, the incident photon is completely reflected, with $T=0$. Thus, tuning the giant atom's size to achieve $\tau$ values satisfying $ \theta \pm \tau\Omega = (2n + 1)\pi$ enables simultaneous control over the number of decoupling points and total reflection points.

	In Fig.~\ref{fig5}(a), with $\tau\gamma_{L1}=13/7$ and $\Delta=\Omega$ satisfying $\theta+\tau\Omega=(2n+1)\pi$, we observe that the transmission reaches unity ($T=1$), indicating complete decoupling between the giant atom and the waveguide in the BEC regime. However, when $\Delta=-\Omega$, the photon is partially reflected, as marked by blue diamonds. In the BUEC regime (i.e., $(\gamma_{L2}, \gamma_{R2})/\gamma_{L1} =(0.25,1)$), we find $T(\Delta=\Omega)=0$ while $T(\Delta=-\Omega) \neq 0$, as marked by yellow dots. 
	In Fig.~\ref{fig5}(b), with $\tau\gamma_{L1}=13/9$ and $\Delta=-\Omega$ satisfying $\theta-\tau\Omega=(2n+1)\pi$, the BEC regime exhibits $T(\Delta=-\Omega)=1$ with waveguide decoupling, while $T(\Delta=\Omega) \neq 1$ (blue diamonds). The BUEC regime shows $T(\Delta=-\Omega)=0$ while $T(\Delta=\Omega) \neq 0$ (yellow dots).
	When $\tau\gamma_{L1}=5/3$ satisfies both  $\theta+\tau\Omega=(2n_{1}+1)\pi$ and $\theta-\tau\Omega=(2n_{2}+1)\pi$, the system exhibits contrasting behaviors at $\Delta=\pm\Omega$: In the BEC regime, the giant atom decouples from the waveguide, resulting in perfect photon transmission (blue diamonds); In the BUEC regime, complete reflection occurs at these frequencies (yellow points), as demonstrated in Fig.~\ref{fig5}(c).

	\section{\label{Sec:4} Conclusion}
	
	This study investigates the single-photon scattering characteristics of a single $\Lambda$-type giant atom chirally coupled to a 1D waveguide at two points. By analyzing two chiral coupling modes—bidirectional even coupling (BEC) and bidirectional uneven coupling (BUEC), we find that the symmetry of the scattering spectrum depends on the phase difference between the two coupling points: when the phase difference is $n\pi$, the spectrum remains symmetric; otherwise, the symmetry is broken. 
	In contrast to the two-level giant atom, the external driving field induces splitting in the scattering spectrum, transforming the single-dip structure into a double-dip profile with perfect transmission ($T=1$) at $\Delta=0$. The frequency separation between the two dips increases with the driving field strength.
	Further research demonstrates that by tuning chiral coupling, incident photons can be flexibly switched between perfect transmission and total reflection. In the Markovian limit, the bidirectional even coupling system always exhibits perfect transmission at a phase of $\theta=(2n+1)\pi$, independent of the driving field. In contrast, in the non-Markovian regime, increasing the distance between coupling points enhances oscillatory behavior in the scattering spectrum. Additionally, the size of the giant atom can modulate the number of decoupling points and perfect reflection points, providing new degrees of freedom for controlling single-photon routing. 

	
	\begin{acknowledgments}
		This work was supported by NSFC Grants
		No.12247105, No.92365209, No.12421005, XJ-Lab
		Key Project (23XJ02001), and the Science $\And $  Technology Department of Hunan Provincial Program
		(2023ZJ1010).
	\end{acknowledgments}
	

\end{document}